# Towards A Census of Earth-mass Exo-planets with Gravitational Microlensing

## White Paper submission to the ESA Exo-Planet Roadmap Advisory Team

J.P. Beaulieu[1], E. Kerins[2], S. Mao[2], D. Bennett[3], A. Cassan[4], S. Dieters[1,5], B.S. Gaudi[6], A. Gould[6], V. Batista[1], R. Bender[7,8], S. Brillant[9], K. Cook[10], C. Coutures[1], D. Dominis-Prester[11], J. Donatowicz[12], P. Fouqué[13], E. Grebel[4], J. Greenhill[5], D. Heyrovsky[14], K. Horne[15], D. Kubas[9], J.B. Marquette[1], J. Menzies[16], N.J. Rattenbury[2], I. Ribas[17], K. Sahu[18], Y. Tsapras[19], A. Udalski[20], C Vinter[21]

[1] Institut d'Astrophysique de Paris, 98 bis boulevard Arago, 75014 Paris, France
[2] Jodrell Bank Centre for Astrophysics, University of Manchester, Manchester, M13 9PL, UK
[3] Department of Physics, University of Notre Dame, IN 46556, USA
[4] Astronomisches Rechen-Institut, Heidelberg University, Monchhofstr. 12-14, 69120 Heidelberg, Germany
[5] University of Tasmania, School of Maths and Physics, Private bag 37, GPO Hobart, Tasmania 7001, Australia
[6] Departement of Astronomy, Ohio State University, Columbus OH 43210, USA
[7] Universitaets-Sternwarte der, Ludwig-Maximilians-Universitaet, Scheinerstr. 1, D-81679 Muenchen, Germany
[8] Max-Planck-Institut fuer Extraterrestrische Physik, Giessenbachstrasse, D-85748 Garching, Germany
[9] European Southern Observatory, Casilla 19001, Vitacura 19, Santiago, Chile
[10] Lawrence Livermore National Laboratory, IGPP, PO Box 808, Livermore, CA 94551, USA
[11] Physics Department, Faculty of Arts and Sciences, University of Rijeka, 51000 Rijeka, Croatia
[12] Technical University of Vienna, Dept. of Computing, Wiedner Hauptstrasse 10, Vienna, Austria
[13] Observatoire Midi-Pyrenees, UMR 5572, 14 Av. Ed. Belin, 31400 Toulouse, France
[14] Institute of Theoretical Physics, Charles University, V Holesovickach 2, 18000 Prague, Czech Republic
[15] SUPA, University of St Andrews, School of Physics & Astronomy, North Haugh, St Andrews, KY16 9SS, UK
[16] South African Astronomical Observatory, PO Box 9 Observatory7935, South Africa
[17] Institut de Ci`encies de l'Espai (CSIC-IEEC), Campus UAB, 08193 Bellaterra, Spain
[18] Space Telescope Science Institute, 3700 San Martin Drive, Baltimore, MD 21218, USA
[19] Las Cumbres Observatory, 6740 Cortona Dr. Suite 102, Santa Barbara, CA, USA
[20] Warsaw University Observatory, Al Ujazdowskie 4, 00-478 Warszawa, Poland
[21] Niels Bohr Institute, Astronomical Observatory, Juliane Maries Vej 30, 2100 Copenhagen, Denmark

## Abstract

Thirteen exo-planets have been discovered using the gravitational microlensing technique (out of which 7 have been published). These planets already demonstrate that super-Earths (with mass up to ~10 Earth masses) beyond the snow line are common and multiple planet systems are not rare. In this White Paper we introduce the basic concepts of the gravitational microlensing technique, summarise the current mode of discovery and outline future steps towards a complete census of planets including Earth-mass planets. In the near-term (over the next 5 years) we advocate a strategy of automated follow-up with existing and upgraded telescopes which will significantly increase the current planet detection efficiency. In the medium 5-10 year term, we envision an international network of wide-field 2m class telescopes to discover Earth-mass and free-floating exo-planets. In the long (10-15 year) term, we strongly advocate a space microlensing telescope which, when combined with Kepler, will provide a complete census of planets down to Earth mass at almost all separations. Such a survey could be undertaken as a science programme on Euclid, a dark energy probe with a wide-field imager which has been proposed to ESA's Cosmic Vision Programme.

## 1. Introduction

Whether we are alone in the Universe is one of the most profound questions mankind has ever posed. Remarkably, science is now beginning to answer this question. In the last fifteen years, astronomers have found close to 300 exo-planets, including some in systems that resemble our very own solar system (Gaudi et al. 2008). These discoveries have already challenged and revolutionised our theory of planet formation and dynamical evolution.

Several different methods have been used to discover exo-planets, including radial velocity, stellar transits, and gravitational microlensing. Exo-planet detection via gravitational microlensing is a relatively new method (Mao & Paczyński 1991, Gould & Loeb 1992) and is based on Einstein's theory of general relativity. So far 13 exo-planets have been discovered with this method. While this number is relatively modest compared with that discovered by the radial velocity method, microlensing probes part of the parameter space (host separation vs. planet mass) not accessible in the medium term to other methods (see Figure 3). The mass distribution of microlensing exo-planets has already revealed that cold super-Earths (at or beyond the "snow line" and with a mass of around 5–15 Earth masses) appear to be common (Beaulieu et al. 2006, Gould et al. 2007). Microlensing is currently capable of detecting cool planets of super-Earth mass from the ground and, with a network of wide-field telescopes strategically located around the world, could detect cool Earth mass to super-Earth mass planets; free-floating planets can also be detected when they are young and just ejected during the formation of the planetary systems (see Section 3). Such free-floating planets are impossible to detect with other methods. Microlensing is sensitive to planets around the most common types of star, while other methods typically target solar-like stars. It is therefore an independent and complementary detection method for aiding a comprehensive understanding of the planet formation process.

The planets discovered by microlensing are usually a few kpc away and orbiting faint stars, and thus too distant for direct imaging or radial velocity followup. The strength of gravitational microlensing is to provide statistics of exo-planets in part of the parameter space not accessible to other methods (Figure 3). Microlensing probes mostly exo-planets outside the snow line, where the favoured core accretion theory of planet formation predicts a larger number of low-mass exo-planets (Ida & Lin 2005). The statistics provided by microlensing will enable a critical test of the core accretion model.

Exo-planets probed by microlensing are much further away than those probed with other methods. They provide an interesting comparison sample with nearby exo-planets, and allow us to study the extrasolar population throughout the Galaxy. In particular, the host stars with exo-planets appear to have higher metallicity (e.g. Fischer & Valenti 2005). The metallicity of stars in the Galactic bulge range over -1.5<[Fe/H]<0.5, with the average metallicity being slightly less than solar (e.g. Fulbright et al. 2005). However, the frequency of planets increases quadratically with increasing metallicity. Taking these two factors into account, it has been estimated that the frequency of planets in the Galactic bulge should be similar to that of the solar neighbourhood, which is supported by SWEEPS observations of transiting planets in the Galactic bulge (Sahu et al. 2006). It will be interesting to see whether similar results hold for microlensing exo-planets.

This White Paper first introduces the basic principle of gravitational microlensing (Section 2), and highlights its advantages and complementarity to other methods. In Section 3, we outline future strategies in the near (within 5 years), medium (5-10 years) and long term (10-15 years), with the ultimate goal of achieving a full census of Earth-like planets. In Section 4, we briefly summarise our recommendations. The conclusions and the roadmap are similar to the white papers submitted to the exoplanet Task force (Bennett et al. 2007, Gould et al. 2007), the exoplanet forum (Gaudi et al. 2008), and the JDEM request for information (Bennett et al. 2008b).

## 2. Basic microlensing principles

The physical basis of microlensing is the deflection of light rays by a massive body. A distant source star is temporarily magnified by the gravitational potential of an intervening star (the lens) passing near the line of sight, with an impact parameter smaller than the Einstein ring radius $R_E$, a quantity which depends on the mass of the lens, and the geometry of the alignment. For a source star in the Bulge, with a 0.3 $M_\odot$ lens, $R_E \sim 2$ AU, the projected angular Einstein ring radius is ~1 mas, and the time to transit $R_E$ is typically 20-30 days, but can be in the range of 5-100 days. The lensing magnification is determined by the degree of alignment of the lens and source stars (see Figure 1).

A planetary companion to the lens star will induce a perturbation to the microlensing light curve scaling with the square root of the planet's mass, lasting typically a few hours (for an Earth) to a few days (for a Jupiter). Hence, planets can be discovered by dense photometric sampling of ongoing microlensing events (Mao & Paczyński 1991, Gould & Loeb 1992).

The inverse problem, finding the properties of the lensing system (planet/star mass ratio, star-planet projected separation) from an observed light curve, is a complex non-linear one within a wide parameter space. In general, model distributions for the spatial mass density of the Milky Way, the velocities of potential lens and source stars, and a mass function of the lens stars are required in order to derive probability distributions for the masses of the planet and the lens star, their distance, as well as the orbital radius and period of the planet by means of Bayesian analysis. With complementary high angular resolution observations, either on board HST or with adaptive optics, it is possible to get additional constraints to the parameters of the system, and determine masses to better than 20% by directly measuring the lens and source relative proper motion (Bennett, Anderson & Gaudi 2007; Dong et al. 2008, Bennett et al. 2006). A space-based microlensing survey can provide the planet mass, projected separation from the host, host star mass and its distance from the observer for most events using this method.

## 2.1 Overview of key microlensing survey and follow-up groups

The observational challenge is several-fold: first there is a need to detect a large number of ongoing microlensing events towards the Galactic Bulge. This is achieved by frequent monitoring of the Bulge by wide field imagers. Currently the Polish OGLE collaboration has a 1.3m telescope equipped with a 0.4 deg$^2$ camera observing from Las Campanas. The Japanese/New Zealand MOA collaboration has a 2 deg$^2$ camera on a 1.8m telescope in New Zealand. In 2007, 800 ongoing microlensing events were detected. Once OGLE upgrades its camera to 1.4 deg$^2$ in 2009 (OGLE IV phase), we can expect to have around 1000 microlensing events per year. Moreover, these two groups are observing repeatedly several square degrees in the Galactic Bulge in order to have dense sampling (up to one image every 20 minutes). And last but not least, both have a specific procedure to be able to analyze quickly the ongoing microlensing events, and trigger denser observations on anomalous microlensing events, or high magnification ones (with higher planet detection efficiencies).

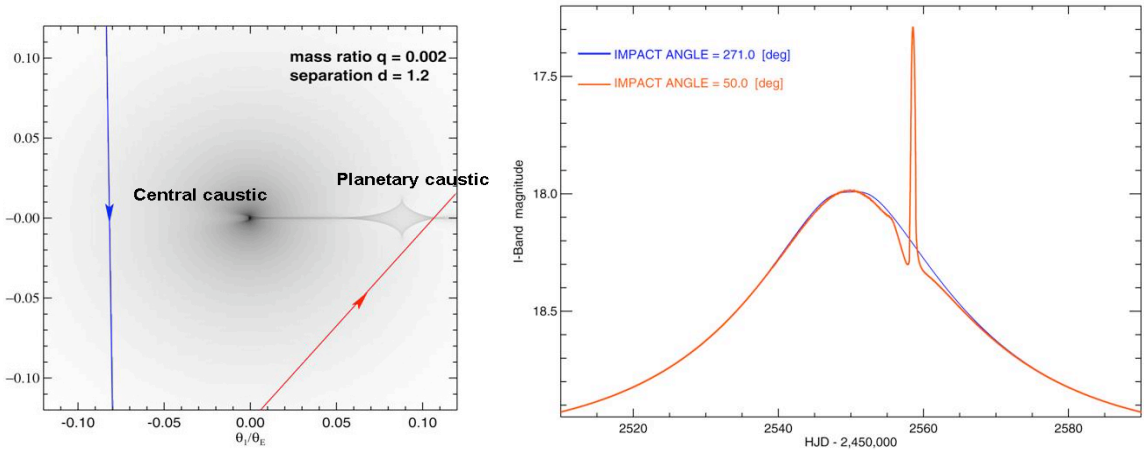

**Figure 1:** The left panel shows the caustic structure of a star/planet lens, with two possible trajectories of a source star. The right panel shows the corresponding observed light curves. Hitting the planetary caustic or passing close to it induces a short-lived but clearly detectable photometric signal. While the impact parameter is the same for the two trajectories shown, the presence of an observable deviation due to the planet strongly depends on the orientation of the source trajectory relative to the star-planet axis. Strong signals can result if the source trajectory intercepts a caustic, whereas there are configurations for which the light curve (e.g. the left trajectory) is essentially identical to that of a star with no planet.

These survey telescopes are supplemented with a fleet of telescopes (from 35cm amateur telescopes to 2m-class robotic, and also targeted observations with VLT, KECK or HST) to achieve round-the-clock monitoring and detect real time deviations in the photometric signal. The two current main networks are the PLANET/RoboNET (Beaulieu et al. 2007) and the µFUN (Gould et al. 2006) collaborations; in the near-term, LCOGT, MONET and other proposed networks will vastly increase ground-based robotic follow-up capabilities.

The four consortia (MOA, OGLE, PLANET/RoboNET and µFUN) are all following the same philosophy to reach the same objective: all the data are reduced in real time, and made available to the community in order to increase the efficiency of planet discovery. Modelling of microlensing events is also made available to the community, in order to properly determine the nature of the ongoing phenomenon. Some microlensing events have been observed by up to 20 telescopes, providing continuous coverage and cross checks between telescopes, data reduction and model fitting methods.

## 2.2 Planet discoveries from Survey+Follow-up model

Since 2003, 13 planets have been detected by microlensing out of which 7 have already been published (Bond et al. 2004, Udalski et al. 2005, Beaulieu et al. 2006, Gould et al. 2006, Gaudi et al. 2008, Bennett et al. 2008a, Dong et al. 2008). Two planets have been discovered by the surveys (MOA and OGLE) independently of the follow up networks, whilst most have been confirmed with additional follow-up observations. Most have been discovered in real time while the anomaly was still ongoing, and denser observations have been obtained by additional telescopes. Moreover, most have been discovered in high magnification events perturbed by the central caustic (or a resonant caustic). At least one system contains multiple planets (Gaudi et al. 2008) which highlights the fact that central caustic crossing events (see Figure 2) are particularly sensitive to the presence of multiple planets in lensing systems.

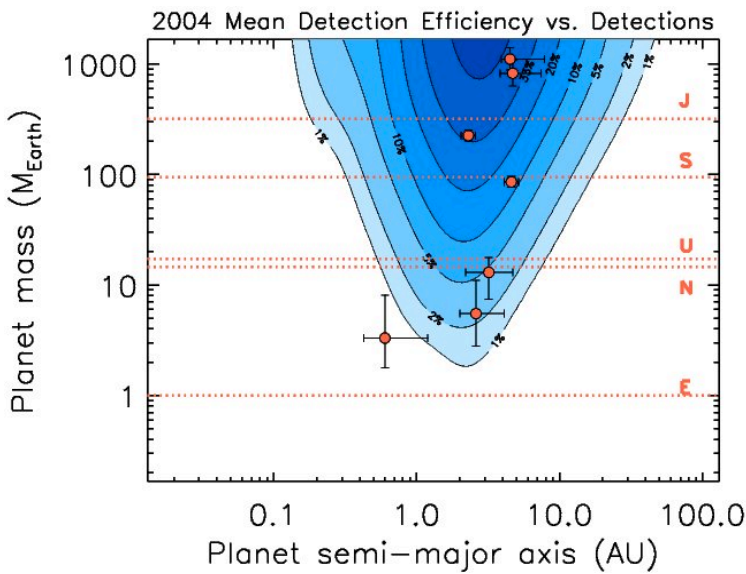

**Figure 2:** Average follow-up detection efficiency of planets as a function of mass and orbital separation (assuming circular orbits) in all events with planetary sensitivity monitored by PLANET in 2004 (Cassan et al. 2008). Seven published planets detected by microlensing are marked. For Jupiter-mass planets, the detection efficiency reaches 50%, while it decreases only with the square root of the planet mass until the detection of planets is further suppressed by the finite size of the source stars for planets with a few Earth masses. Nevertheless, the detection efficiency still remains a few per cent for planets below 10 Earth masses. The masses of several solar system planets are indicated by the dotted horizontal lines.

## 2.3 Early implications of these discoveries

Although 13 planets is a small number compared to the harvest made by the radial velocity method, these detections (out of which 7 have been published) have already provided important constraints on planet formation theories. We are probing regions of parameter space that are in the near/medium term inaccessible to other methods, having a maximum sensitivity beyond the snow line (see Figure 3).

The publication of three cool, "super-Earth" planets suggest that these planets are common (Beaulieu et al. 2006, Gould et al. 2006, Bennett et al. 2008a). The detection of a Jupiter/Saturn analogue also suggests that solar system analogues are probably not rare (Gaudi et al. 2008).

## 3. Strategies for the next 15 years

Microlensing surveys undertaken from both the ground and from space offer the best prospects of completing a full census of planets down to an Earth mass over separation scales exceeding 1 AU. The technique provides a nice complement to Kepler, a NASA mission to be launched in 2009 which will provide a comparable census for separations below 1 AU using the transit method (see Figure 3). The achievement of this goal requires not just a European-wide coordinated effort, but a high degree of international coordination. The microlensing community, both in Europe and internationally, has a firm track record of cooperation and data sharing.

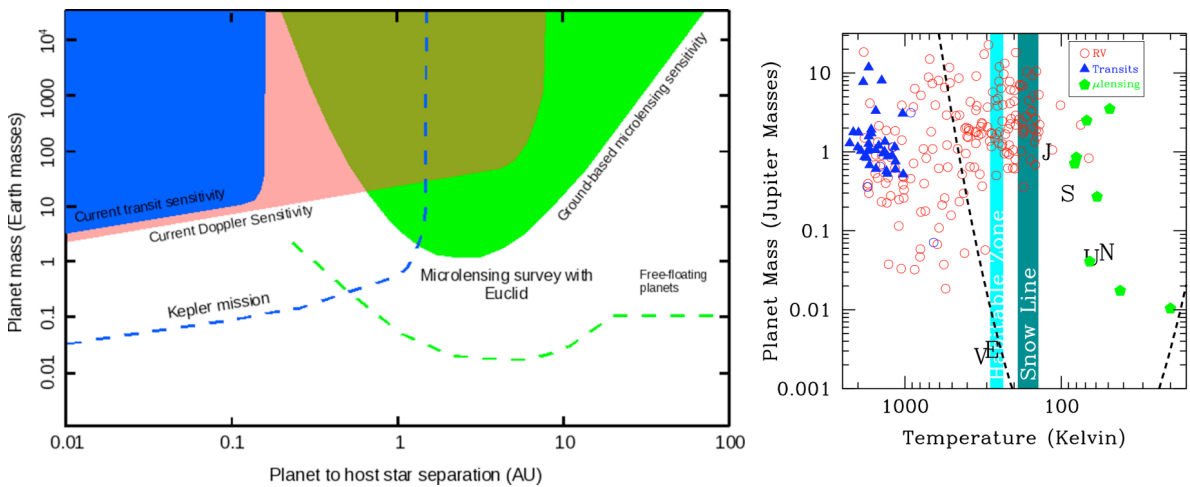

**Figure 3:** (*Left*) The discovery space of exo-planets via the Doppler (red), transit (blue) and microlensing (green) methods. The combination of Kepler and space-based microlensing would enable a complete census of planets down to Earth mass across virtually the entire range of host star separations. (*Right*) The effective temperature versus mass for discovered exo-planets, including published microlensing detections. The habitability zone and the snow line are also marked. Microlensing sensitivity to planets beyond the snow line provides good complementarity with the transit and Doppler methods. The positions of Solar System planets are indicated by their initials. The dashed line roughly indicates the region where a dedicated space-based microlensing survey would detect >10 planets if every main-sequence star had one of the given mass and equilibrium temperature.

Over the next 15 years exo-planet detection with microlensing can be expanded in three distinct stages, beginning with more automated follow-up (over the next 5 years), through to the use of wide-field telescope networks (with a full network expected to operate within a 5-10 year time-frame), and ultimately through to a space-based microlensing survey (in a 10-15 year time-frame). Below we discuss the main factors affecting the efficiency of current networks, and then go on to describe each of the three stages, and their associated milestones.

*3.1 Current limitations*

Despite the high degree of international cooperation the efficiency of current follow-up technique can be significantly improved (Figure 2 is close to the current situation). Current limitations and critical points are:

1. It is critical to obtain high quality photometry in real time at the different telescopes. There has been a major improvement from going from point-spread-function fitting photometric pipelines, to image subtraction based pipelines (Alard 2000), but there is still room for improvement.
2. We need to identify early on the handful of high magnification microlensing events that have the best sensitivity to exo-planets among the 800-1000 alerted microlensing events every year.
3. We need to identify deviations from point source point lens microlensing early and disentangle spurious photometric variation from real anomalies.
4. To date, only point lens microlensing model are fitted real time automatically. There is a need for automatic binary lens with finite source effects fitting online.
5. From the ground most main sequence bulge stars are unresolved and so when they are microlensed the magnified flux is strongly blended with unmagnified flux from neighbouring stars. This severely limits sensitivity to light-curve deviations due to planets and also tends to increase modelling degeneracies. Events involving bright bulge giant stars are less affected by blending. However, their large size typically produces finite source size effects, which wash out microlensing perturbations for planets below about a few Earth masses. The sensitivity of ground-based surveys to planets below a few Earth masses is therefore limited. However, in very high magnification events, a main sequence bulge star (with smaller size) is resolved and therefore has sensitivity to lower mass planets. To date, the lowest mass planet detected is ~3.3 $M_\oplus$.
6. Alerted events can be followed from the ground only when observing conditions permit. Gaps in datasets due to bad weather is one of the limitations of ground-based exo-planet microlensing, since the exo-planet deviations themselves are short-lived.
7. Since many events are ongoing simultaneously a decision must be made on how intensively to follow the ones with the highest in detection efficiency, and collect data on promising events to be able to predict their future behaviour. There is a need for coordination and spread of the resources in the best possible way.
8. The data quality from follow-up networks is heterogeneous, involving telescopes with different band-passes observing from sites with a range of seeing quality. This impacts upon the photometric quality of the light-curves and also upon the complexity of microlensing modelling, especially if wavelength-dependent effects, such as finite source size and blending, are evident in the data.
9. Searching the parameter space of binary microlensing models is currently computationally expensive and time consuming. Effects such as finite source size, relative parallax motion of the lens and source, or the orbital motion of the planetary companion can also modify the microlensing magnification, in which case the modelling requires additional parameters.

*3.2 The near-term: automated follow-up*

The observing efficiency of follow-up surveys, can be improved by optimizing the scheduling of observations in order to maximize the potential for anomaly detection. There are two approaches to this. The first is to concentrate on high-magnification events, which are particularly sensitive to planets (Griest & Safizadeh 1998). The majority of microlens planet detections to date came from high magnification events. A second approach is to allocate observing time across all ongoing events in a way which attempts to optimize the planet discovery efficiency. The international PLANET/RoboNet/HOLMES group has been advancing this approach using human-assisted and robotic telescope networks. Incoming photometry from all the monitored events is updated, refined and re-fitted as it is collected. Regular human inspection by the observers and members of the collaboration, together with anomaly-detection algorithms are then employed to determine whether there is evidence of an ongoing anomaly in any of the events. If an ongoing anomaly is suspected the algorithm can determine the most efficient observing strategy to confirm and characterise it. This produces a new observing schedule that, when embedded within an intelligent agent system (such as eSTAR, see e.g. Snodgrass et al 2008), can allow an entire telescope network to operate seamlessly in gathering further data. The new observations are then re-processed and the cycle repeats, producing an observing strategy that is both adaptive and economic with observing time. One critical bottleneck to the exercise is to collect and reduce real time high quality photometry.

In order to fully exploit automated observations it is desirable to be able to interpret the light-curve in real time, perhaps allowing future observing strategies which are designed to distinguish between the future predicted behaviour of currently acceptable theoretical light-curve fits. ARTEMiS (Dominik et al. 2008) is one implementation of such a system. To be able to fit a full binary lensing model in real-time to continuously updated data in order to inform future observing schedules is highly demanding. The fitting algorithm must be able to incorporate data from all available sources, respond intelligently to data gaps, and be able to identify and deal with inevitable systematic errors in the photometry due, e.g., to poor observing conditions.

A fully intelligent and responsive follow-up network should significantly increase the efficiency of planet detection and, importantly, will make it easier to assess the underlying completeness of such surveys to planet discovery. It is important for undertaking the first census of the cold planet population. The infrastructure for making rapid automated studies of anomalies in ongoing events will also be required for larger-scale ground- or space-based surveys.

**Milestones:**

**A) An optimised planetary microlens follow-up network, including feedback from fully-automated real-time modelling.**

**B) The first census of the cold planet population, involving planets of Neptune- to super-Earth mass (few $M_\oplus$ to 20 $M_\oplus$) with host star separations around 2 AU.**

**C) Under highly favourable conditions, sensitivity to planets close to Earth mass with host separations around 2 AU.**

*3.3 The medium-term: wide-field telescope networks*

Even a fully optimised follow-up network cannot expect to detect all ongoing planetary events. Currently around 800 microlensing events per year are alerted by the OGLE and MOA survey teams but, whilst this is a large number, there is likely a comparable number of events which are not identified in real time but which can be recovered in a more complete off-line analysis. Some of these missed events will have potentially detectable planetary signals. Furthermore, planet formation theories often predict a population of free-floating planets that escape their host systems during planet formation and evolution. Microlensing events involving free-floating planets will usually be too short to be detected by the main survey teams and so are not targeted by follow-up teams. Evidence of a free-floating planet population would hold important clues for planet formation theories and microlensing is the only available technique able to probe this channel.

The next generation of microlensing surveys will involve wide-field 1-2m class telescopes which are able to survey tens of millions of stars with a cadence of 10-20 minutes. Round the clock monitoring will be possible by establishing a dispersed network of such telescopes. The advantages of such a network are that the planet discovery efficiency is no longer dependent on the ability to alert on ongoing events; all events within the survey area can be searched off-line for the presence of planets. The high survey cadence also makes it possible to undertake the first census of the free-floating planet population. These surveys will typically identify several thousand new microlensing events per year and so the planet discovery rate should increase by at least an order of magnitude. This will allow a complete census to be made for planets down to ~10 $M_\oplus$, below which finite source effects limit the ground-based planet detection efficiency.

The first steps towards such a network are already being undertaken. The MOA survey has recently upgraded its telescope system (MOA-2) to a 1.8m aperture mirror and a 2 deg$^2$ camera operating at Mt John in New Zealand. This is effectively the first node of the future wide-field network. The OGLE survey is soon making an upgrade from OGLE-III to OGLE-IV. The new OGLE-IV survey will operate from Chile with first light expected in 2009. It will have a factor of ~4 increase in data flow over OGLE-III using a 1.4 deg$^2$ camera. This will effectively create the second node of the wide-field network. The deployment of a full network of three or four nodes will likely take around five years. A Korean node looks likely to be sited in South Africa, and possibly a Chinese-led node at Dome A in Antarctica, which is potentially one of the best optical sites in the World.

Whilst the hardware for the wide-field network is already taking shape, much of the development that will be required to make such a network operate successfully must take place at the software level. The huge volume of light-curves will require fully automated algorithms for fitting and interpreting large numbers of events, though these will not need to operate in real time. A lot of experience will have been gained from the automated follow-up era, although the computational demands of the wide-field network will be higher still.

**Milestones:**

A) **Complete census of the cold planet population down to ~10 $M_\oplus$ with host separations above 1.5 AU.**

B) **The first census of the free-floating planet population.**

C) **Sensitivity to planets close to Earth mass with host separations around 2 AU.**

*3.4 The longer-term: a space-based microlensing survey*

Even with a fully-fledged wide-field network, there will remain significant limits to the capabilities of planet detection from the ground. These are emphasized by points 5 and 6 in Section 3.1. Only by monitoring main sequence stars can we overcome the finite-source size effects that otherwise usually wash out perturbing signals from planets below 10 $M_\oplus$. Weather also remains a limiting factor to light-curve coverage. The only way to obtain uninterrupted surveillance of large numbers of well resolved main sequence stars is to conduct a survey from space, where we escape atmospheric and weather limitations.

Detailed simulations (Bennett & Rhie 2002) show that a dedicated space-based microlensing mission can detect planets down to 0.1 $M_\oplus$ and moons (Bennett et al. 2007). A space-based survey allows all analogues of our own Solar System (measured relative to host mass), with the sole exception of Mercury, to be detected. It can also detect free-floating planets down to Earth mass as well as measure tens of thousands of planetary transit events. An important additional advantage highlighted in Section 2 is that usually one can obtain a complete lens solution for microlensing events observed from space (Bennett, Anderson & Gaudi 2007). This is rarely true for ground-based microlensing without additional space-based follow-up, which may be difficult to do for all planet candidates detected by the wide-field network discussed in Section 3.3. A complete lens solution for large numbers of planetary events would, for example, allow the distribution of planet masses to be studied as a function of distance from the observer.

Whilst a dedicated space-based microlensing survey could extend the census down to 0.1 $M_\oplus$, a cost-effective survey down to Earth masses could be achieved as a secondary science activity aboard a dark energy space mission, which has very strict telescope and camera requirements that are ideally suited to microlensing. The Euclid dark energy survey is a candidate for a medium-class mission within the ESA Cosmic Vision Programme and, if selected by ESA, would launch around 2017. Euclid was recently ranked very highly by the ASTRONET Infrastructure Roadmap

Working Group (Bode et al 2008). The Euclid baseline specification is a 1.2m telescope with a CCD optical imager, a near-IR array and a near-IR multi-object slit spectrometric channel (Laureijs 2008). Its primary science objective is to constrain the nature of dark energy through both weak lensing and baryon acoustic oscillations, involving a weak lensing analysis of around 3 billion galaxies over 20,000 deg$^2$ of the sky and the measurement of spectroscopic redshifts for a subset of around 100 million galaxies. The requirements for weak lensing are high spatial resolution, a very stable point spread function and a wide-field imager. These same requirements also make Euclid an ideal facility for a microlensing exo-planet survey. Microlensing is among the potential secondary science objectives identified in the Euclid Science Requirements Document (Laureijs 2008). With a few months observations per year during the primary science mission, together with additional time after the primary science, a *J*-band microlensing survey using Euclid would be capable of completing a census of planets down to 1 $M_\oplus$ over host separations exceeding 1 AU.

**Milestones:**

- **A) A complete census of planets down to Earth mass with separations exceeding 1 AU.**
- **B) Complementary coverage to Kepler of the planet discovery space.**
- **C) Potential sensitivity to planets down to 0.1 $M_\oplus$, including all Solar System analogues except for Mercury.**
- **D) Complete lens solutions for most planet events, allowing direct measurements of the planet and host masses, projected separation and distance from the observer.**

4. **Summary and International Context**

We have described a roadmap for exo-planet detection with the microlensing technique which can enable a complete census of planets down to Earth masses to be achieved over the next 15 years. Microlensing is a powerful method for the detection of extra-solar planets, as recognized recently in the Report of the ExoPlanet Task Force by the NSF Astronomy and Astrophysics Advisory Committee in the US. Europe has played and continues to play a leading role in this field with leadership of OGLE, PLANET, HOLMES and RoboNet.

Over the next five years we anticipate automated follow-up surveys to provide: a) an optimised planetary microlens follow-up network, including feedback from fully-automated real-time modelling; b) the first census of the cold planet population, involving planets above 10 $M_\oplus$ with host star separations around 2 AU; and c) under highly favourable conditions, sensitivity to planets close to Earth mass with host separations around 2 AU. Currently European researchers are leading or playing key roles in programmes such as PLANET, HOLMES, RoboNet and ARTEMiS which are pushing forward this approach.

On a five to ten-year time-frame wide-field telescope networks will be deployed to provide: a) a census of the cold planet population down to 10 $M_\oplus$ with host separations above 1.5 AU; b) the first census of the free-floating planet population; c) sensitivity to planets close to Earth mass with host separations around 2 AU. Currently one node of such a network has been deployed by the Japanese/New Zealand MOA team, and another similar telescope will expand its capability and enter a new phase in 2009 operated by the Polish OGLE team. There are concrete plans for a similar Korean telescope to be sited in South Africa, and tentative plans for a European led telescope at Dome C or a Chinese led telescope based at Dome A in Antarctica.

Ultimately, a comprehensive census of cold planets down to Earth masses requires a space-based microlensing survey. On a ten to fifteen-year time-frame a space-based survey could launch and provide: a) a complete census of planets down to Earth mass with separations exceeding 1 AU; b) complementary coverage to Kepler of the planet discovery space; c) sensitivity to planets down to 0.1 $M_\oplus$, including all Solar System analogues except for Mercury; d) complete lens solutions for most planet events, allowing direct measurements of the planet and host masses, and distance from the observer. There is a great potential opportunity for such a survey to be carried out as a full mission such as Microlensing Planet Finder (Bennett et al. 2007) or as a secondary science objective for a dark energy mission (Bennett et al. 2008b, Refregier et al. 2008). Euclid, which is a candidate for a medium class mission within ESA's Cosmic Vision Programme, could launch around 2017 if selected.